\documentclass[a4paper,11pt]{article}
\usepackage{latexsym}
\usepackage{amssymb}
\usepackage{amsmath}
\usepackage{epstopdf}
\usepackage{cite}
\usepackage{authblk}
\usepackage{graphicx}
\usepackage{subcaption}
\usepackage{epsfig}
\usepackage{float}
\begin{document}

\title{Symmetry breaking, and the effect of matter density on neutrino oscillation}
\author{H. Mohseni Sadjadi \thanks{mohsenisad@ut.ac.ir}}
\author{A. P.  Khosravi Karchi\thanks{amir.khosravi@ut.ac.ir}}
\affil{Department of Physics, University of Tehran}
\renewcommand\Authands{ and }
\maketitle
\begin{abstract}
 A proposal for the neutrino mass, based on neutrino-scalar field interaction, is introduced. The scalar field is also non-minimally coupled to the Ricci scalar, and hence relates the neutrino mass to the matter density.  In a dense region, the scalar field obeys the $Z_2$ symmetry, and the neutrino is massless. In a dilute region, the $Z_2$ symmetry breaks and neutrino acquires mass from the non-vanishing expectation value of the scalar field.  We consider this scenario in the framework of a spherical dense object whose outside is a dilute region.  In this background, we study the neutrino flavors oscillation, along with the consequences of the theory on oscillation length and MSW effect. This preliminary model may shed some lights on the existing anomalies within the neutrino data, concerning the different oscillating behavior of the neutrinos in regions with different densities.

\end{abstract}

\section{Introduction}
It has been almost 60 years since the idea of massive neutrinos and their oscillations has been proposed \cite{osc1,osc2,osc3,osc4,osc5,osc6,osc7,osc8}. While the idea is well established and experimentally confirmed, some anomalies and data are still out there,  waiting for proper explanations. Among these, the 2$\sigma$ level difference between $\Delta{m}^2$ values measured for the solar and Earth experiments\cite{Anomalies-cpt,Anomalies-cpt2}, and the detection of $\overline{\nu}_e$ excess in the $\overline{\nu}_{\mu}$ beam by LSND\cite{Anomalies-LSND} and MicroBooNE\cite{micro,micro1}, are the ones which motivate people to construct new models or to complete the old ones. Though there are many models which can remove some of the above-mentioned problems\cite{Proposed,Proposed1,Proposed2,Proposed3}, they involve CPT and Lorentz violation, or predict the existence of a new sterile neutrino not yet been shown up in the experiments and which is heavily constrained by the theoretical and experimental considerations\cite{sterile,sterile1,sterile2}. It is possible that both of the above-mentioned anomalies are due to different neutrino masses in various environments. The main motivation for the idea of matter density dependent neutrino mass, is the difference between environmental densities in which the neutrino oscillations have taken place.  As for the Sun and the Earth, we have $\frac{\rho^{mean}_{S}}{\rho^{mean}_{E}} \simeq 0.2$, and in many situations the observed neutrinos have passed through mediums with drastically different densities. The influence of matter density on neutrino oscillation via the weak interaction, is considered in the MSW (Mikheyev–Smirnov–Wolfenstein)effect\cite{matter effects,matter effects1,matter effects2} within which neutrino physics is affected by the local electron density. The effect of some non-standard interactions on neutrinos oscillation has also been studied in the literature \cite{ns1,ns2}.

A MSW like effect due to the neutrino interaction with an exotic scalar field, is discussed in \cite{gold}, where the scalar field gives an effective mass to the neutrinos. In \cite{string}, based on neutrino interaction with a long rage scalar field in the context of string theory, a model for neutrinos oscillation is proposed. The possible role of scalar fields as the dark energy, and the comparability of the dark energy density scale and the neutrino mass splitting, inspired people to develop models comprising neutrino-scalar field interactions \cite{first,first1}. Neutrino-scalar field coupling is also studied in mass varying neutrino models, in which the transition of neutrinos from relativistic to non-relativistic phase, by affecting the effective quintessence potential, gives rise to the Universe's late time acceleration, \cite{MaVaN,sad}.

In this paper we relate the neutrino mass, and consequently the neutrino oscillation, to environmental matter density, via its interaction with a scalar field which is non-minimally coupled to the Ricci scalar \cite{on}. Such non-minimal couplings are vastly studied in the literature in the context of inflation and late time acceleration. This coupling, implicitly, provides a relation between neutrino oscillation and matter density. We assume a $\mathbb{Z}_2$ symmetry for the scalar field section, which is spontaneously broken when the matter density decreases, this scenario somehow resembles the symmetron model discussed in  \cite{Screening,Screening1,mat,mat1,sad2,sad4,mot}. The interaction between the scalar field and the neutrinos and its coupling to the curvature, can lead to different results from the previous studies of the neutrino-scalar interaction\cite{string,Our,Our1,Our2,Our4}. In section $2$, we introduce the model and derive analytic solutions for the scalar field and then study the neutrinos oscillation.  We also point out to the influence of the new interaction on the MSW effect. Finally, in the conclusion, we conclude and discuss our results through a numerical example. We adopt the natural units $\hbar=c=1$ and the metric signature $(-,+,+,+)$.

\section{Non-minimally coupled scalar field, neutrino mass,  and symmetry breaking }
Our model is specified by the action
\begin{eqnarray}\label{1}
S&=&\int d^4x {\sqrt{-g}}[\frac{M_{P}^2R}{2}-\frac{1}{2}g_{\mu \nu} \partial^\mu{\phi} \partial^\nu{\phi}-V(\phi)-\epsilon\frac{R}{2} \phi^2+\kappa_i\phi^2\overline{\nu_{i}}\nu_{i}\nonumber \\
&&-i \overline{\nu_{i}}\gamma^\alpha D_\alpha \nu_{i}+\mathcal{L}(\Psi, g_{\mu \nu})],
\end{eqnarray}
where $\phi$ is the scalar field whose potential is $V(\phi)$, and $M_P\simeq 2.4\times 10^{18}GeV$ is the reduced Planck mass. $\nu_i$ denotes ith neutrino whose mass depends on $\phi$.   $\mathcal{L}(\Psi, g_{\mu \nu})$ is the matter Lagrangian density (comprising ordinary and dark matter). $\phi$ is also non minimally coupled to the Ricci scalar.
So the scalar field interacts with non-relativistic matter via the nonminimal coupling to the Ricci scalar, and also with the neutrino via its mass term. In this way we plan to relate the neutrino mass to the matter density and the symmetry breaking procedure.

The non-minimal coupling term may be related to the requirement that the theory be renormalizable in first loop corrections.  Such a coupling has also been employed to study the inflation and the late time cosmic acceleration. Introducing new scalar degrees of freedom in the gravitation equations may violate the equivalence principle \cite{eq}. In regions where the scalar field is highly screened, this violation may not be detected by local gravitational tests. The source of the scalar field in (\ref{1}), is the the trace of the energy momentum tensor, so its universal coupling to matter components implies the weak equivalence principle (WEP) \cite{Screening1}. However as we will discuss, in the end of this subsection, WEP can be violated for astrophysical extended objects in screening models \cite{eq,eq1}.

The neutrino mass term, i.e. $\kappa_i\phi^2\overline{\nu_{i}}\nu_{i}$ is $\phi$ dependent. Although this term may be considered as a phenomenological interaction(like other phenomenological interactions between the exotic scalar field with dark matter, sterile neutrinos and so on, considered in the literature), but may also has root in more fundamental theories. In \cite{MaVaN,Xi}, based on seesaw mechanism,  a neutrino mass section in the Lagrangian is considered:
\begin{equation}
\mathcal{L}_{n.m.}=\mathcal{M}_{ij}l_{Li}\nu_{Rj}H+\frac{1}{2}M_{ji}(\phi)\bar{\nu}_{Rj}^{c}\nu_{Ri}+h.c.,
\end{equation}
where $\nu_R$ is the right handed sterile neutrino whose mass depends on the exotic scalar field $\phi$: $M_{ji}(\phi)=\frac{M_{ji}}{f(\phi)}$.  $l_L$ and $H$ denote left handed lepton and Higgs doublets. $<H>\mathcal{M}_{ij}$ is the Dirac mass. By integrating the sterile neutrinos, a $\phi$ dependent mass, proportional to $f(\phi)$, will be generated for other neutrinos through a dimension-5 operator (for details see \cite{Xi}).  In our model $f(\phi)=\phi^2$, which respects the $Z_2$ symmetry.
By another way in \cite{cl},  it was explained how this mass term
can arise from a higher-dimensional operator in supergravity models. Regarding the naturalness of the model, radiative instability of the scalar degrees of freedom is a well-known result of Coleman-Weinberg mechanism\cite{phd}. The $\phi-Higgs$ coupling can produce loop corrections to $\left<\phi\right>$\cite{fab}, hence making the model unnatural unless a new mechanism, like supersymmetry, cancels the contributions towards the Higgs vacuum expectation value. The scalar-neutrino interaction term also introduces quadratically divergent corrections towards the scalar mass which leads to upper bounds on the scale of new physics associated with neutrinos\cite{MaVaN,Bu}. The details of such bounds are model dependent and are beyond the scope of this paper.

By varying the action with respect to the metric, we obtain the Einstein's equation
\begin{eqnarray}\label{2}
M_{P}^2G_{\mu \nu}&=&T_{\mu \nu}^{(m)}+T_{\mu \nu}^{(\nu)}+\epsilon(g_{\mu \nu}\Box-\nabla_\mu \nabla_\nu+G_{\mu \nu})\phi^2+\partial_\mu \phi \partial_ \nu\phi\nonumber \\
&&-\frac{1}{2}g_{\mu \nu}\partial_\sigma \phi \partial^\sigma \phi-g_{\mu \nu}V(\phi)+g_{\mu \nu} \kappa_{i} \phi^2 \overline{\nu_{i}} \nu_{i}.
\end{eqnarray}
By taking the trace of (\ref{2}), we arrive at
\begin{equation}\label{3}
R=-\frac{T^{(m)}+T^{(\phi)}_{eff.}}{M_P^2},
\end{equation}
where $R$ is the Ricci scalar, $T^{(m)}=T^{(m)\mu}_\mu$, and we have defined $T^{(\phi)}_{eff.}$ as
\begin{equation}\label{4}
T^{(\phi)}_{eff.}:=\epsilon(3\Box-R)\phi^2-\partial^\mu \phi \partial_\mu \phi-4V(\phi)+4\kappa_i\bar{\nu_i}\nu_i \phi^2.
\end{equation}

In the following, we assume that $T^{m}\gg T^{\phi}$, such that $R\simeq  \frac{\rho^{(m)}}{M_P^2}$,  i.e. the main contribution in the curvature is coming from the matter whose energy density is much larger than its pressure.

Varying the action with respect to the scalar field, gives
\begin{equation}\label{5}
\Box \phi -V_{,\phi}-\frac{\epsilon}{M_P^2}  \rho^{(m)}\phi+2\kappa_i\bar{\nu_i}\nu_i \phi=0.
\end{equation}
 The metric is assumed to be mainly determined by a spherical source with mass $M$, with a constant density $\rho$, and radius $r_0$, whose inside $\frac{\rho r^2}{M_P^2}\ll 1$ holds, while, for the outside, we have  $\frac{M}{M_P^2r}\ll 1$.  Hence it is safe to take $\Box$  the same as its form in the Newtonian limit \cite{Screening1}.

We consider the Higgs like potential
\begin{equation}\label{6}
V(\phi)=V_0-\frac{\mu^2}{2} \phi^2+\frac{\lambda}{4} \phi^4,
\end{equation}
where $V_0$ is a constant. Therefore, provided that $4\kappa_i\bar{\nu_i}\nu_i \ll \mu^2$, (\ref{5}) reduces to
\begin{equation}\label{7}
\frac{d^2\phi}{dr^2}+\frac{2}{r}\frac{d\phi}{dr}= (\frac{\epsilon}{M_P^2}  \rho^{(m)}-\mu^2)\phi+\lambda \phi^3.
\end{equation}

We define the critical density, $\rho_c$, by
\begin{equation}\label{8}
\rho_c=\frac{M_P^2\mu^2}{\epsilon}
\end{equation}
For  $\rho^{(m)}\geq \rho_c$, the $Z_2$ symmetric effective potential defined by
\begin{equation}\label{9}
V_{eff.}=V_0+\frac{1}{2}(\frac{\epsilon}{M_P^2} \rho^{(m)}-\mu^2)\phi^2+\frac{\lambda}{4}\phi^4,
\end{equation}
 has a minimum at $\phi=0$,  $V_{eff.}(\phi=0)=V_0$.  For $\rho^{(m)}<\rho_c$, minimum occurs at one of the points $\phi_b=\pm  \sqrt{\frac{\mu^2}{\lambda}(1-\frac{ \rho^{(m)}}{\rho_{c}})}$. If in the whole region, the matter density was larger that the critical density, the solution of (\ref{7}) would become $\phi=0$, implying a zero mass for neutrinos.

Now let us assume that the region is divided into two parts with different densities: $\rho^{(m)}(r<r_0)=\rho_{in}>\rho_c,\,\,$ and $\rho^{(m)}(r>r_0)=\rho_{out}^{(m)}<\rho_c$. For this configuration the continuous solution of (\ref{7}), around the minimum, which is regular at $r=0$ is
\begin{equation}\label{10}
\phi(r)= \begin{cases}
\frac{A}{r} sinh(m_{in} r) & r\leq r_0 \\
\phi_b+\frac{B}{r} e^{-m_{out} r} & r_0\leq r,
\end{cases}
\end{equation}
where the mass is given by $m^2=\frac{d^2 V_{eff.}}{d\phi^2}$, evaluated at the minimum of the effective potential. Hence $m_{in}=\sqrt{\left(\frac{\rho_{in}}{\rho_c}-1\right)}\mu$, and
$m_{out}=\sqrt{2 \left(1-\frac{\rho_{out}}{\rho_c}\right)}\mu$. $\phi_b$ is the asymptotic value of $\phi$, and if $\rho_{out}\ll \rho_c$, then $\phi_b^2=\frac{\mu^2}{\lambda}$. Continuity of $\phi$ and its derivative at $r=r_0$ implies
\begin{equation}\label{11}
\begin{split}
\text{\small{A}} &=\frac{\phi_{b}(1+m_{out} r_0)}{m_{in} cosh(m_{in} r_0)+m_{out} sinh(m_{in}r_0)} \\
\text{\small{B}} &=-e^{m_{out} r_{0}} \phi_{b}\frac{m_{in} r_0-tanh(m_{in} r_0)}{m_{in}+m_{out}tanh(m_{in}r_0) }
\end{split}
\end{equation}

By assuming $m_{in}r_0\gg 1 $ and $m_{out}r_0\ll 1$, which result in the screening inside the source object and a long
range force outside it, the solution of the scalar field can be simplified to:
\begin{equation}\label{12}
\centering
\phi(r)=\begin{cases}
     \frac{\phi_{b}}{m_{in} cosh(m_{in} r_0)} \frac{sinh(m_{in} r)}{r}  & r\textless r_0 \\
      \phi_{b}-\phi_{b} \frac{m_{in} r_0 - tanh(m_{in} r_0)}{m_{in}r}e^{-m_{out}(r-r_0)} & r_0\textless r
\end{cases}
\end{equation}

This shows that the neutrino mass evolves from  $\kappa_i\frac{\phi_b}{\cosh(m_{in}r_0)}\ll \kappa_i\phi_b$ near the origin in the dense region,  to $\kappa_i\phi_b$ at  $r\gg r_0$ in the outside rare region. This is due to the symmetry breaking. Note that the change of the matter density changes the profile of the effective potential outside the spherical object.
We remind that if the matter density did not cross the critical density, the neutrino would remain massless.

\begin{figure}[H]
\centering
\begin{subfigure}[b]{0.5\linewidth}
\includegraphics[width=\linewidth]{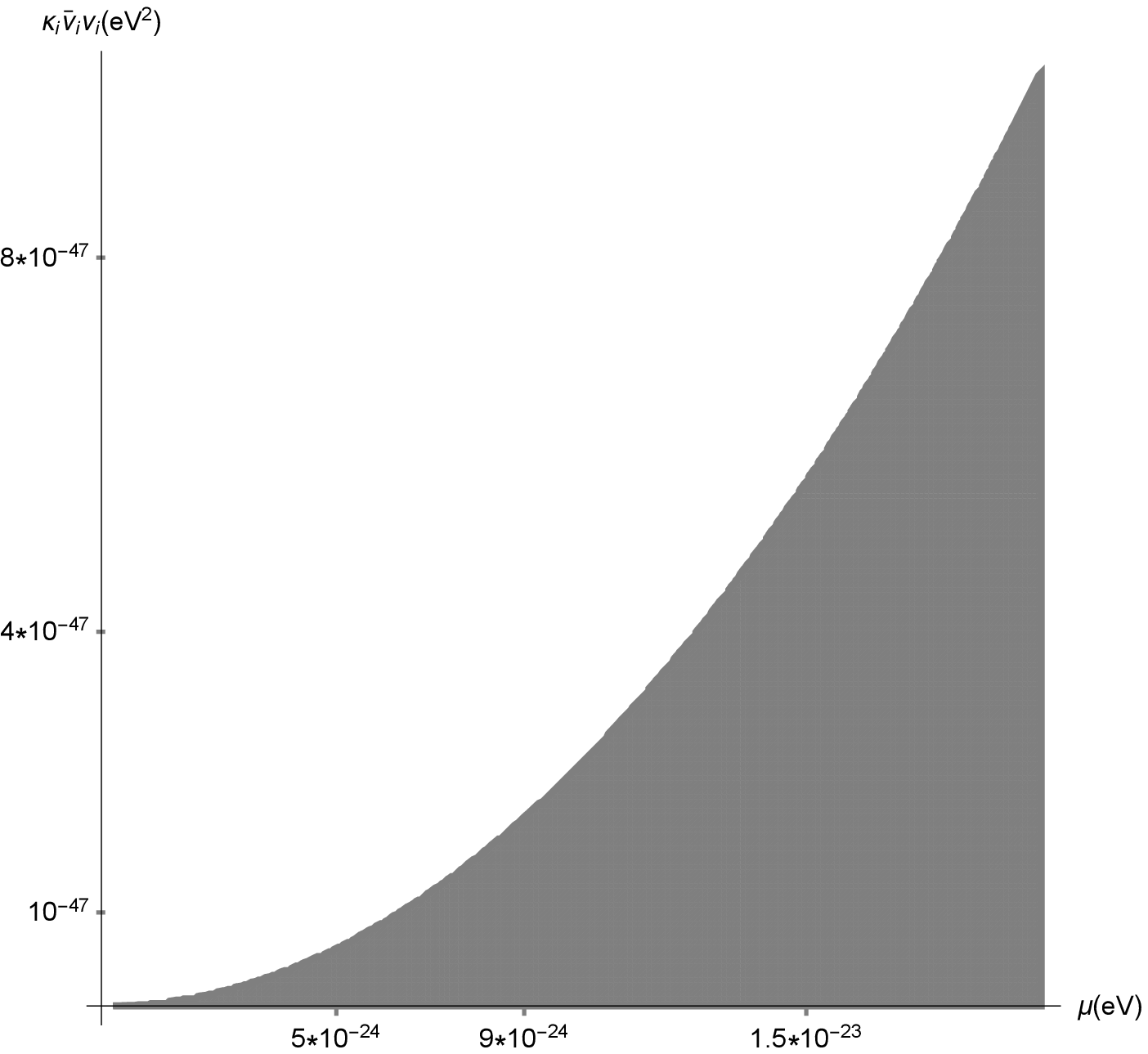}
\end{subfigure}
\begin{subfigure}[b]{0.5\linewidth}
\includegraphics[width=\linewidth]{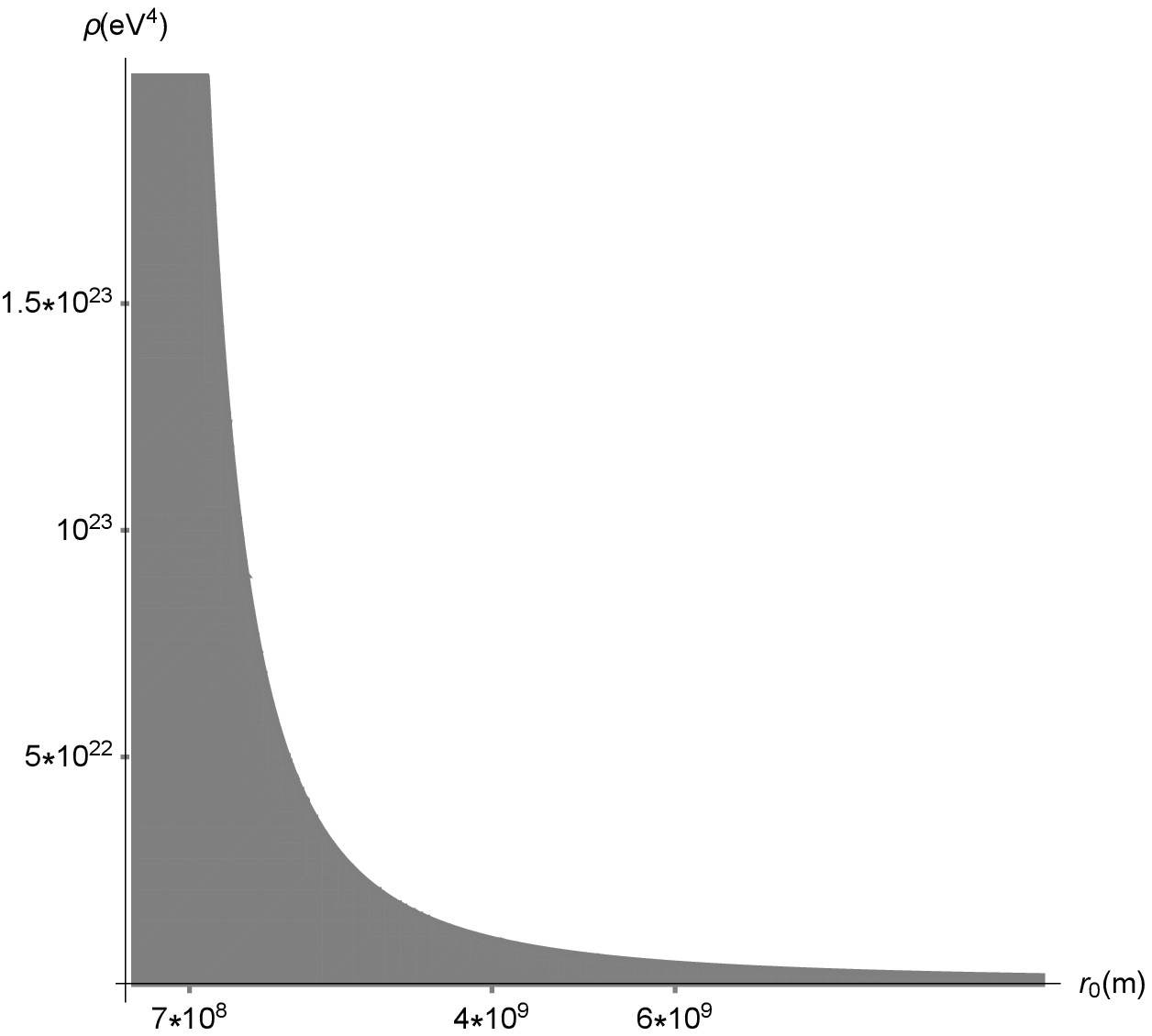}
\end{subfigure}
\caption{Regions allowed by $4 \kappa_i \overline{\nu_i}\nu_i \ll \mu^2$ (up) and $\frac{\rho r_0^2}{M_P^2} \ll 1$ (down).}
\label{fig:1}
\end{figure}

\begin{figure}[H]
\centering\epsfig{file=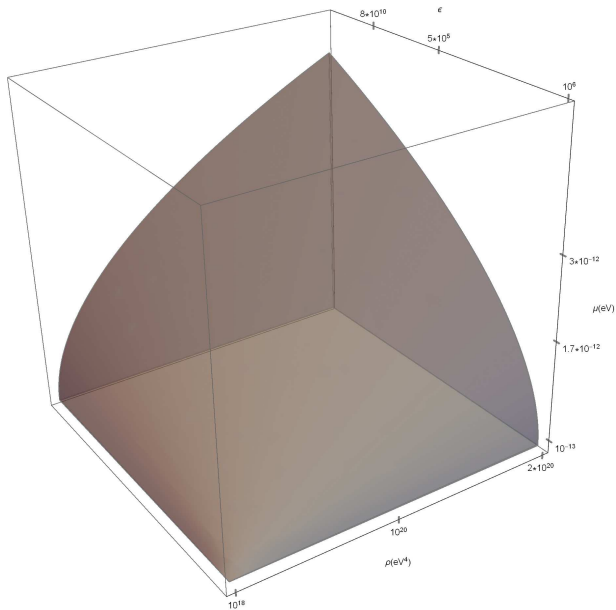,width=8cm}
\caption{ Allowed range of parameters by $\rho_c < \rho$} \label{fig:2}
\end{figure}

In derivation of the solutions, we have required that three main inequalities hold: $4 \kappa_i \overline{\nu_i}\nu_i \ll \mu^2,\,\,\,\frac{\rho r^2}{M_P^2} \ll 1$ and $\rho_c < \rho_{in}$. To know  the allowed regions in the parameter space, we note that $\frac{\rho r_0^2}{M_P^2} \ll 1$ guaranties both $\frac{\rho r^2}{M_P^2} \ll 1$ inside, and $\frac{M}{M_P^2r}\ll 1$ outside the source. The allowed regions are depicted in figures (\ref{fig:1}), and (\ref{fig:2}).

Note that (\ref{5}) implies a new force mediated by the scalar field.  For two test point masses located at $r_1$ and $r_2$, and for a light scalar field, i.e.  $\mu r_i\ll 1$, the ratio of this force to the Newtonian gravitational force is derived as $\frac{F_{\phi}}{F_N}=2\epsilon^2\frac{\phi(r_1)\phi(r_2)}{M_P^2}$ . In dense regions where $\rho>\rho_c$ we have $<\phi>=0$, implying $F_{\phi}=0$. In regions where $\rho<\rho_c$, but $\epsilon^2\phi(r_1)\phi(r_2)\ll {M_P^2}$, this force is negligible with respect to the gravitational force.
Following the steps of  \cite{Screening1}, (by replacing  $\frac{1}{M^2}$ in \cite{Screening1} with $\frac{\epsilon}{M_P^2}$), we find that an extended object with spherically symmetry is screened at distances $r_0\ll r\ll \mu^{-1}$, provided that $6\epsilon \Phi \gg 1$, where $\Phi$ is the gravitational potential at the surface of our source object, e.g. for the sun this gives  $\epsilon >10^6$ \cite{Screening1}.

Before ending this subsection, let us note that in the screening models there is the possibility that WEP be violated in the macroscopic level. This lies on the fact that extended objects, with different internal compositions, fall at different rates in a gravitational field. Typically, one can introduce a scalar charge, $Q \sim M(1-\frac{M(r_s)}{M})$, in which $M$ is the total mass of the source object and $M(r_s)$ is the amount mass embedded inside the screening radius\cite{eq1}. This scalar charge enters the right hand side of equation of motions for objects, and since $r_s$ is different for various objects, i.e. some are totally screened and some are not screened at all etc., it leads to violation of WEP\cite{eq1}. However, in the basic model represented here, we take the source object's densities to be of the step-function form and as a result, $r_s$ is either the radius of the object in which case $Q=0$ or it is zero, which leads to a $Q$ that is independent of internal structure.

\subsection{Neutrino flavor oscillation}

A result of taking  mass for neutrinos is the flavor changing effect. A neutrino of flavor $\alpha$, denoted by $\nu_{\alpha}$, may be written in terms of the mass eigenstates, denoted by $\nu_i$, as
\begin{equation}\label{13}
\nu_\alpha(r_a)=\sum_i U_{\alpha i}\nu_i(r_a)
\end{equation}
where $U$ is the PMNS (Pontecorvo-–Maki–-Nakagawa–-Sakata) mixing matrix\cite{osc4}.  Using $\nu_i(r_a)=e^{-i\int_{r_*}^{r_a}p_\mu dx^\mu} \nu_i(r_*)$, for $|{\vec{p}}|\gg m$, the probability to detect a neutrino of flavor $\beta$ at $r_b$ from a neutrino of flavor $\alpha$ at $r_a$ is
\begin{eqnarray}\label{14}
P_{\alpha \to \beta}&=&\left|\left<\nu_\beta(r_b)|\nu_\alpha(r_a)\right>\right|^2\nonumber \\
&=&\sum_{i,j}U^*_{\beta i}U_{\alpha i}U_{\beta j}U_{\alpha j}^* e^{-i\Phi_{ij}}
\end{eqnarray}
where
\begin{equation}\label{15}
\Phi_{ij}=\int_{r_a}^{r_b} \frac{\Delta m_{ij}^2}{2E_0}dr,
\end{equation}
and $\Delta m_{ij}^2=m_i^2-m_j^2$. (\ref{14}) may be written as
\begin{eqnarray}\label{16}
P_{\alpha \to \beta}&=&\delta_{\alpha \beta}-4\sum_{i>j}Re(U^*_{\beta i}U_{\alpha i}U_{\beta j}U_{\alpha j}^*)\sin^2 (\frac{\Phi_{ij}}{2})\nonumber \\
&+&2\sum_{i>j}Im(U^*_{\beta i}U_{\alpha i}U_{\beta j}U_{\alpha j}^*)\sin (\Phi_{ij})
\end{eqnarray}
For a model with two mass eigenstates and two flavour picture, we have
\[
   U=
  \left[ {\begin{array}{cc}
   \cos\theta & \sin\theta\\
   -\sin\theta & \cos\theta\\
  \end{array} } \right]
\]
and consequently (\ref{16}) takes the simple form
\begin{eqnarray}\label{17}
&&P_{\alpha \to \alpha}=P_{\beta\to \beta}=1-\sin^2(2\theta) \sin^2(\frac{\Phi_{21}}{2})\nonumber \\
&&P_{\alpha \to \beta(\neq \alpha)}=\sin^2(2\theta) \sin^2(\frac{\Phi_{21}}{2})
\end{eqnarray}

Using (\ref{15}) and (\ref{1}), we find out that the oscillation phase from $r_a$ to $r_b$, $\Phi_{ij}$, is determined by
\begin{equation}\label{18}
\Phi_{ij} =\frac{\Delta{\kappa_{ij}}^2}{2E_0} \int_{r_a}^{r_b} {\phi^4 dr}
\end{equation}
where $\Delta\kappa_{ij}^2=\kappa_i^2-\kappa_j^2$. For $(r_a<r_0, \,\, r_b<r_0)$, we find
\begin{equation}\label{19}
\Phi_{ij}=\frac{\Delta \kappa_{ij}^2}{2E_0}\left(I(r_b)-I(r_a)\right)
\end{equation}
 while for $(r_a>r_0, \,\, r_b>r_0)$, we have
\begin{equation}\label{20}
\Phi_{ij}=\frac{\Delta{\kappa_{ij}}^2}{2E_0}\left(J(r_b)-J(r_a)\right)
\end{equation}
and finally, for $(r_a<r_0,\,\,\,r_b>r_0)$ we obtain
\begin{equation}\label{21}
\Phi_{ij}=\frac{\Delta{\kappa_{ij}}^2}{2E_0}\left(J(r_b)-I(r_a)\right),
\end{equation}
where
\begin{eqnarray}\label{22}
I(r)&=& \frac{A^4}{24r^3}\Big(\left( 8m_{in}^2 r^2+4 \right) \cosh \left( 2 m_{in}r \right) +
\left( -8 m_{in}^2 r^2-1 \right) \cosh \left( 4 m_{in}r \right)\nonumber \\
&& -16{\it Shi} \left( 2 m_{in}r \right) m_{in}^3r^3+32{\it Shi} \left( 4m_{in}
r \right) m_{in}^3r^{3}+4 \sinh \left( 2m_{in}r \right) m_{in}r\nonumber \\
&&-2 \sinh
\left( 4 m_{in}r \right) m_{in}r-3\Big),
\end{eqnarray}
and
\begin{eqnarray}\label{23}
J(r)&=&\frac{1}{3r^3}\Big(36{\it Ei}\left( 1,2m_{out}r \right) {\phi_b}^{2}{B}^{2}m_{out}{r}^{3
}-54{\it Ei}\left( 1,3m_{out}r \right) \phi_b B^{3}{m_{out}}^{2}{r}^{3}\nonumber \\
&&+32{\it
Ei} \left( 1,4m_{out}r \right) B^{4}{m_{out}}^{3}{r}^{3} -8B^{4} \left( m_{out}
^2r^2-\frac{1}{4}m_{out}r+\frac{1}{8}\right) {{\rm e}^{-4m_{out}r}}\nonumber \\
&&+
18 \left( m_{out}r-\frac{1}{3} \right) B^{3}\phi_br {{\rm e}^{-3m_{out}r}}
-18 {{\rm e}^{-2m_{out}r}}{\phi_b}^{2}B^{2}{r}^{2}\nonumber \\
&&+3{\phi_b }^{4}{r}^{4}-12{\phi_b}^{3}B{\it Ei} \left( 1,m_{out}r \right) {r}^{3}\Big).
\end{eqnarray}
In the above, the {\it Shi} function and the exponential integral function, {\it Ei}, are given by ${\it Shi}(z)=\int_0^z\frac{\sinh t}{t}dt$ and ${\it Ei}(1,z)=-\int_{-z}^{\infty}\frac{e^{-t}}{t}dt$.
$I(r)$ and $J(r)$ have the following limit values
\begin{equation}\label{24}
\lim I(r)=A^4
\begin{cases}
-\frac{m_{in}^2}{6r}exp(4m_{in}r)  & m_{in}r\gg 1 \\
m_{in}^4r & m_{in}r\ll 1
\end{cases}
\end{equation}
and
\begin{equation}\label{25}
\lim J(r)=
\begin{cases}
\phi_b^4r  & m_{out}r\gg 1 \\
-\frac{B^4}{3r^3} & m_{out}r\ll 1
\end{cases}
\end{equation}

$\Phi_{ij}\propto (r_b-r_a)$ happens only for two situations: If $\{r_a<r_0,\,\,\,r_b<r_0,\,\,m_{in}r_a\ll 1,\,\, m_{in}r_b\ll 1\}$, then we obtain $\Phi_{ij}=\frac{\Delta \kappa_{ij}^2}{2E_0}A^4m_{in}^4\left(r_b-r_a\right)$,  similarly for $\{r_a>r_0,\,\,\,r_b>r_0,\,\,m_{out}r_a\gg 1,\,\,m_{out}r_b\gg 1 \}$ we find $\Phi_{ij}=\frac{\Delta \kappa_{ij}^2}{2E_0}\phi_b^4\left(r_b-r_a\right)$.  The behavior of the phase is different in other cases. For example, for  $\{r_a<r_0,\,\, r_b>r_0,\,\,m_{in}r_a\ll 1 ,\,\, m_{out}r_b\gg 1\}$ we find $\Phi_{ij}=\frac{\Delta \kappa_{ij}^2 }{2E_0}\left( \phi_b^4r_b-m_{in}^4r_a\right)$, and for $\{r_a<r_0,\,\, r_b>r_0,\,\,m_{in}r_a\gg 1,\,\, m_{out}r_b\ll 1\}$ we find $\Phi_{ij}=\frac{\Delta \kappa_{ij}^2 \phi_b^4}{2E_0}\left(\frac{8}{3r_am_{in}^2}\exp(4m_{in}r_a)-\frac{r_0^4}{3r_b^3}\right)$.

Note that the oscillation length $L_0$, defined as the distance at which a complete cycle of oscillation happens  i.e. $\Phi_{ij}=2\pi$, only in situations where $\Phi_{ij}\propto (r_b-r_a)$, is a constant, but generally it is position dependent. E. g. for $\{r_a<r_0,\,\,\,r_b<r_0,\,\,m_{in}r_a\ll 1,\,\, m_{in}r_b\ll 1\}$,
$\Phi_{ij}=2\pi$ gives
\begin{equation}
L_0=\frac{4\pi E_0}{\Delta \kappa_{ij}^2 A^4m_{in}^4},
\end{equation}
while  for $\{r_a<r_0,\,\, r_b>r_0,\,\,m_{in}r_a\ll 1 ,\,\, m_{out}r_b\gg 1\}$, we have
\begin{equation}
L_0=r_b-r_a= \left(\frac{m_{in}^4}{\phi_b^4}-1\right)r_a+
\frac{4\pi E_0}{\Delta \kappa_{ij}^2 \phi_b^4},
\end{equation}
which depends on $r_a$.
\subsection{Neutrino-scalar coupling and MSW effect}
So far we have studied the effect of neutrino-scalar coupling on the neutrino oscillation. But neutrinos (at least active neutrinos) have also weak interaction which may affect their oscillations.  During their travels in a dense medium such as the earth and sun, etc., neutrinos may interact with other particles. The influence of such a no-flavor-changing scattering on neutrino oscillation is discussed in the context of the MSW effect \cite{matter effects2}. Based on the fact that electron-neutrino, $\nu_e$, has an additional charged current interaction with electrons, compared to other lepton-neutrinos, an additional potential term in $\nu_e-\nu_e$ sector is added to the Hamiltonian in the flavor basis. After some manipulation one can obtain
\begin{equation}\label{26}
H=
  \frac{\Delta{m}^2}{4E}
  \begin{bmatrix}
    -\cos{2\theta}+A & \sin{2\theta}  \\
    \sin{2\theta} & \cos{2\theta}-A
\end{bmatrix}
\end{equation}
 where
\begin{equation}\label{27}
A=\pm \frac{2\sqrt{2}G_F N_e E}{\Delta{m}^2}
\end{equation}
 originates from the elastic scattering potential $V_e=\pm \sqrt{2}G_F N_e $, where $G_F$ is the Fermi constant, and $N_e$ is the electron density. we consider only two mass eigenstates and $\Delta m^2=m_1^2-m_2^2$.  In regions of high electron density, $A$ dominates the vacuum Hamiltonian, therefore the diagonal elements become more relevant and we get mixing suppression. For $A \simeq \cos{2\theta}$ a resonant enhanced oscillation is resulted. Finally, in dilute environments, $A$ is negligible and we get the ordinary oscillation.
In our model $\Delta m^2$ is position dependent and from screening mechanism, we expect to have $<\phi>\to 0$ in dense regions where $\rho>\rho_c$. In these regions, if filled with ordinary  matter, we expect to have large electron density such that $A$ becomes dominant in the Hamiltonian and the mixing oscillation is suppressed. In contrary in dilute regions, i.e.  $\rho<\rho_c$, due to the spontaneous symmetry breaking , $<\phi>\neq 0$, and we may have resonance and ordinary oscillation.

\section{Results, discussions and conclusion }
We proposed a simple model in which the local matter density has a crucial effect on neutrino oscillation. This effect is established by a scalar field coupled simultaneously to the Ricci scalar and to the neutrinos (see (\ref{1})). We assumed that the model has a $\mathbb{Z}_2$ symmetry (see (\ref{1}) and (\ref{6})) which, when matter density becomes less than a critical value, spontaneously breaks. This means that $\left<\phi\right>\simeq 0$, inside a dense region, while $\left<\phi\right>=\phi_b\neq 0$ in a dilute medium. We considered a high density spherical object with a rare environment outside.  The analytical scalar field  solution for this configuration was obtained (see(\ref{10})). This solution shows that the mass of the neutrino evolves from zero, deep inside the object, to an asymptotic value far away. Using the solution (\ref{10}) , we obtained the neutrino oscillation phase (see(\ref{18}-\ref{21})). In our model, the oscillation length is not generally a constant and is position dependent (see discussion after (\ref{25})). In the last part we briefly pointed out to the possible influence of scalar field-neutrino coupling on MSW effect (see(\ref{26})) and showed that in the dense region as the neutrino mass tends to zero, the mixing oscillation reduces while in rare regions we may have resonance and ordinary oscillations.

To get more intuition about the results, we illustrate our model with a simple numerical example. We pick the numeric values $\mu^2=10^{-24} {eV}^2, \lambda=10^{-36}$,  $\rho_{in}=2 \times 10^{20} {eV}^4$, $m_{out}\simeq \sqrt{2}\mu$, $r_0=7 \times 10^8 m$, $\Delta\kappa_{12}^2=8.8 \times 10^{-29} {eV}^{-2}$, and $\epsilon=10^{15}$. The chosen number for $\epsilon$ is consistent with the limit coming from the screening models (see the notes before subsection 2.1). These numbers imply $m_{in}=4 \times 10^{-10}eV$ and $m_{out}=1.4 \times 10^{-12}eV$. $\rho$ and $r_0$ are intentionally set close to the Sun's average density and radius respectively\cite{sun}. The chosen $\epsilon$ gives $\rho_{c} \simeq 10^{15} {eV}^4=9.35\times 10^{-7}g/{cm^3}$. In our formalism, $\rho_c$ must be much less than the interior density and much larger than the outside density.  Our choice is consistent with the limits for interplanetary matter density in the solar system obtained in \cite{dsolar}: e.g. at the earth's orbital,  $\rho_m<1.4\times 10^{-19}gr/cm^3\ll \rho_c$ .

\begin{figure}[H]
\centering
\begin{subfigure}{0.5\linewidth}
\includegraphics[width=\linewidth]{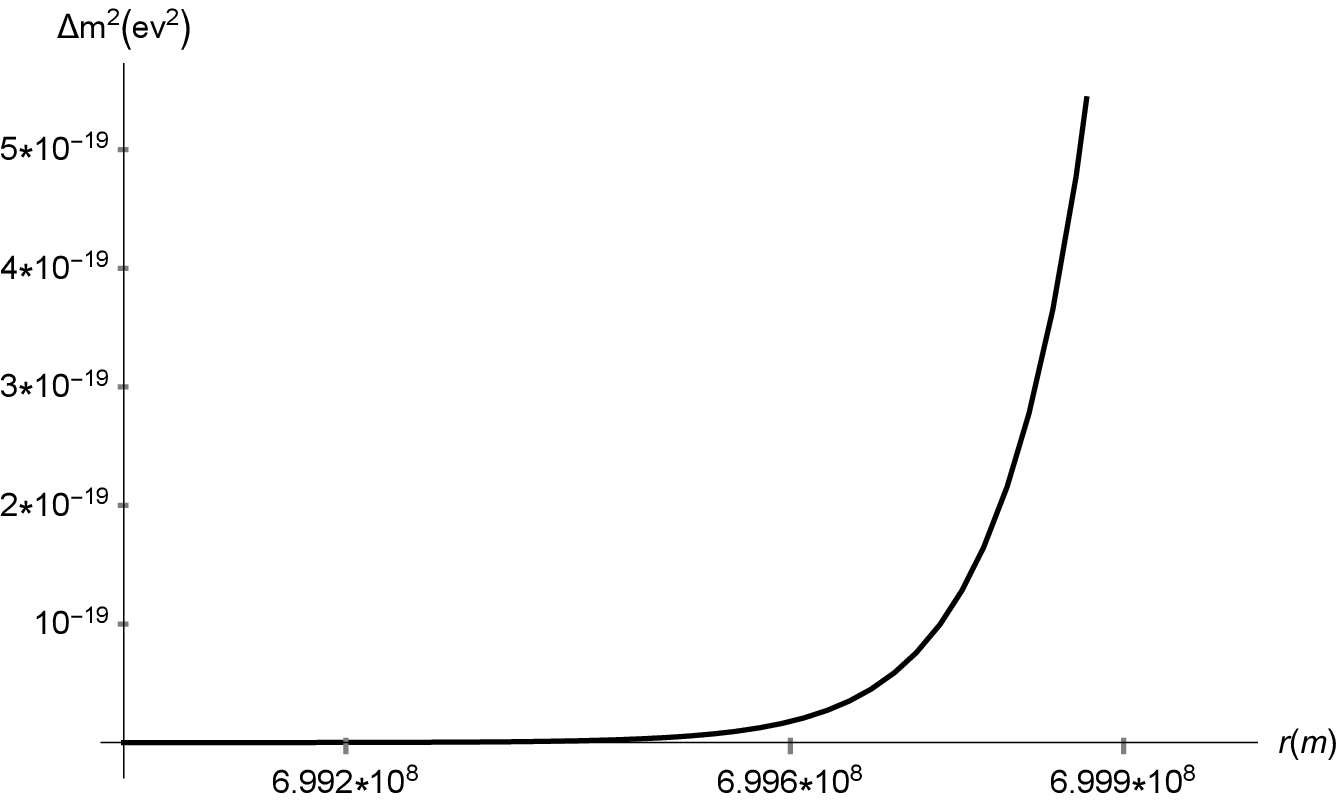}
\end{subfigure}
\begin{subfigure}{0.5\linewidth}
\includegraphics[width=\linewidth]{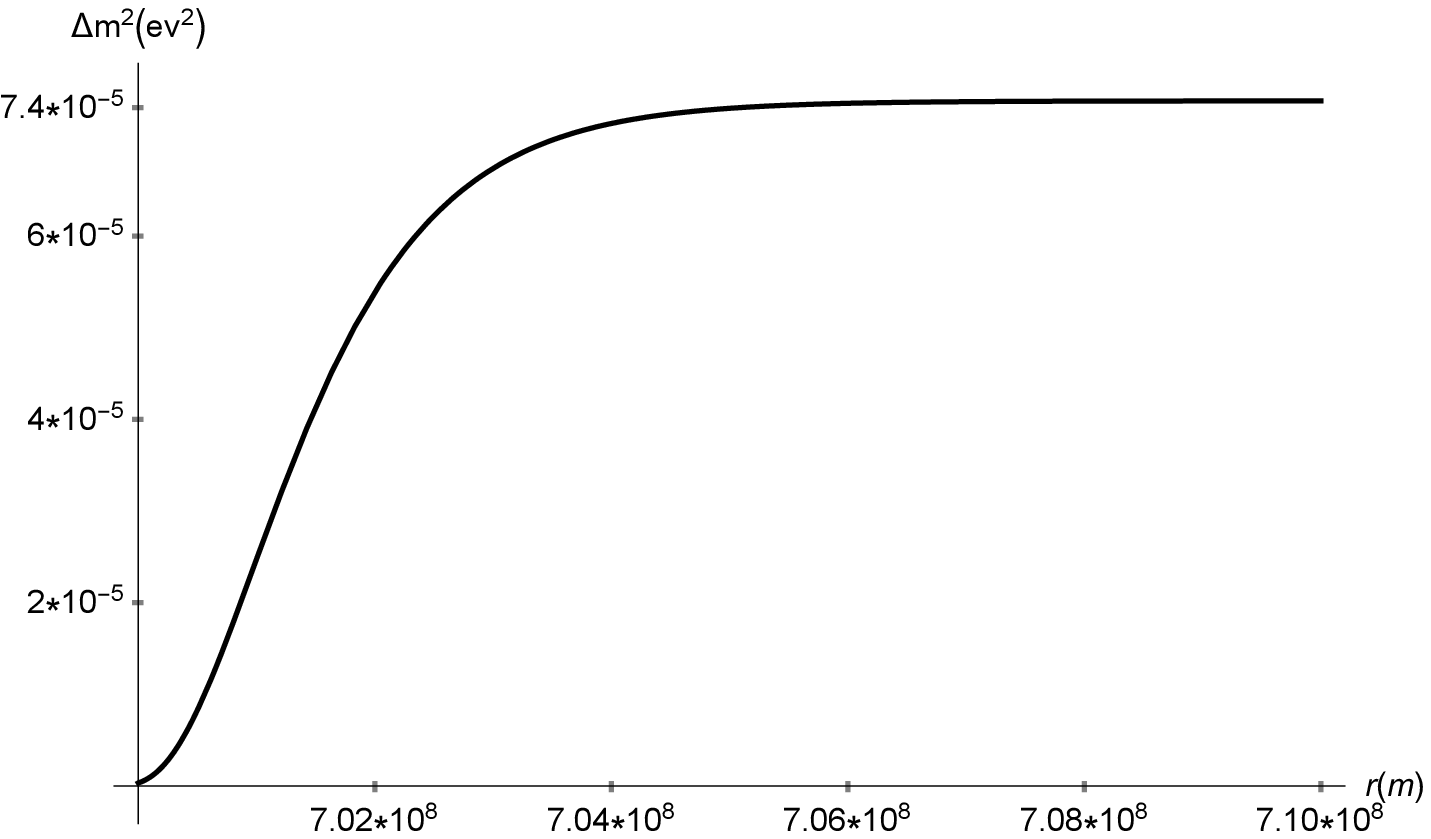}
\end{subfigure}
\caption{$\Delta{m}^2$ as a function of distance inside(up) and outside(down) of a spherical object with radius $r_0=7\times 10^8m$. Note that $\frac{1}{m_{in}} \simeq 3 \times 10^5 m$ and $\frac{1}{m_{out}}\simeq 1.2 \times 10^{6} m$.}
\label{fig:3}
\end{figure}

\begin{figure}[H]
\centering
\begin{subfigure}{0.5\linewidth}
\includegraphics[width=\linewidth]{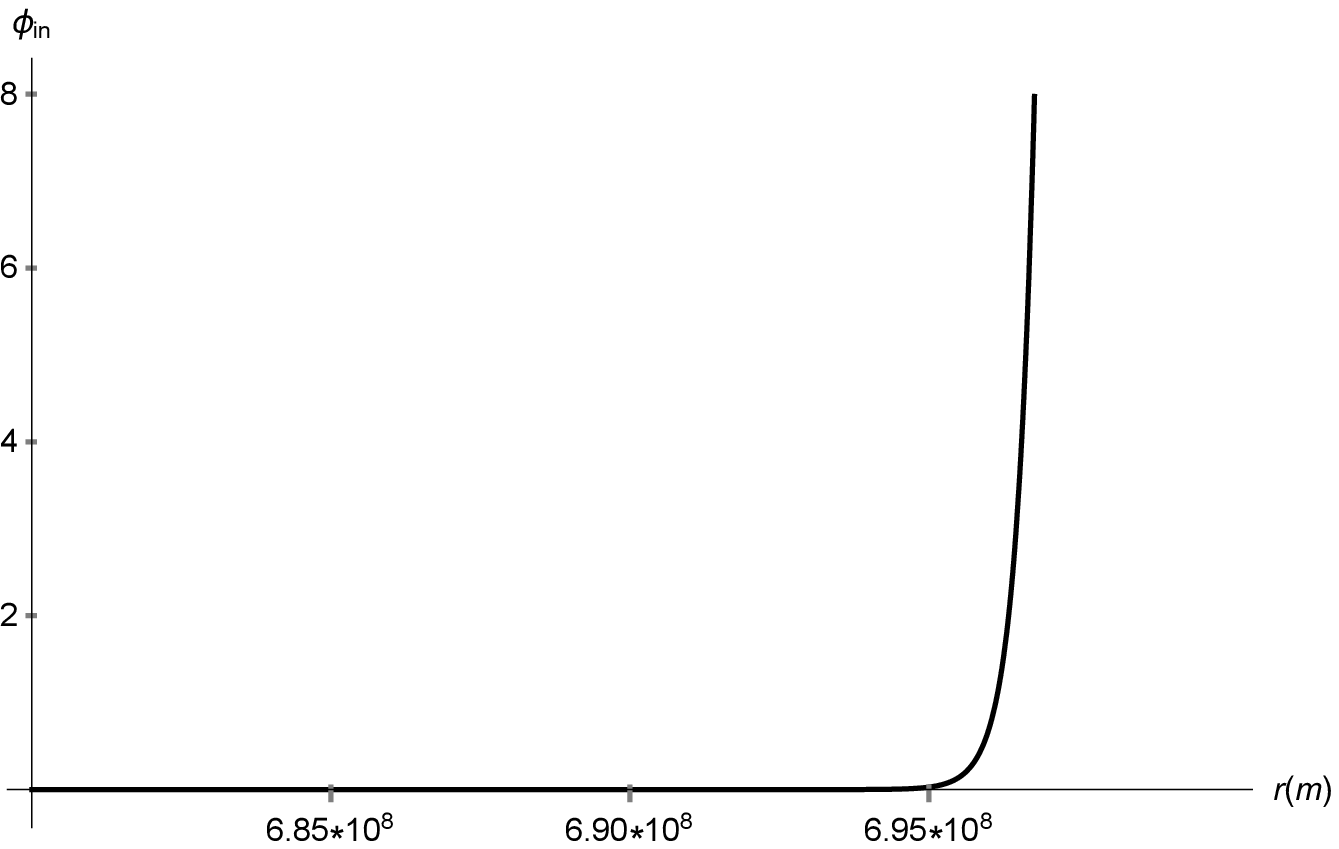}
\end{subfigure}
\begin{subfigure}{0.5\linewidth}
\includegraphics[width=\linewidth]{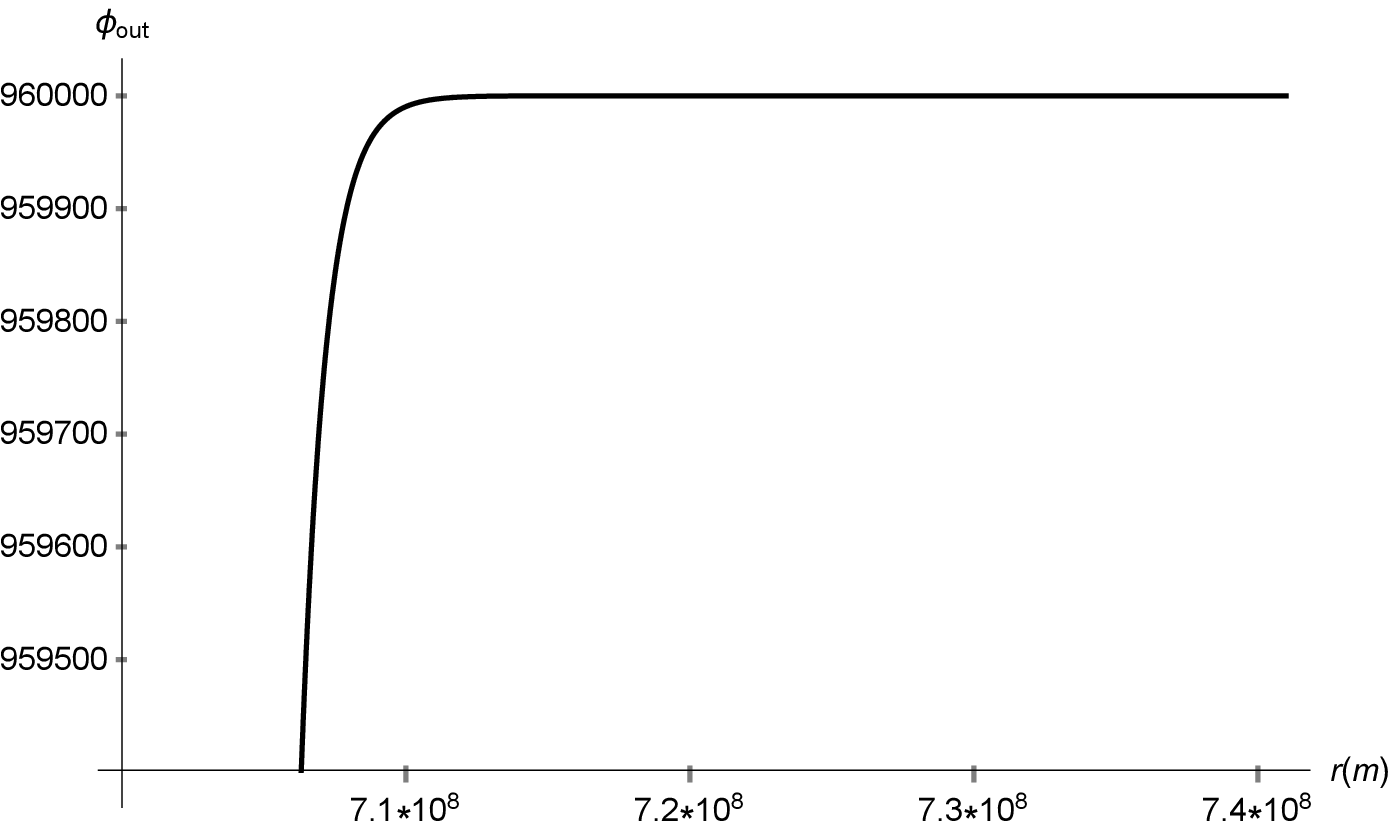}
\end{subfigure}
\caption{The scalar field in units of $eV$, inside(up) and outside(down).}
\label{fig:4}
\end{figure}

We depict $\Delta{m}^2$ and $\phi(r)$ for a typical solar neutrino energy scale of $2 MeV$ in figures (\ref{fig:3}), and (\ref{fig:4}) respectively. Figure (\ref{fig:3}) shows that $\Delta{m}^2$ takes extremely small values as we move closer to the center, with the maximum value  $3 \times 10^{-26} {eV}^2$ right at $r=r_0$, hence the oscillations inside the object are extremely small. Moving further apart, $\Delta{m}^2$ reaches a limiting value $\Delta{m}^2= \Delta\kappa_{12}^2\frac{\mu^4}{\lambda^2}=7.4 \times 10^{-5} {eV}^2$. Regarding the scalar field, it takes a negligible value deep inside the object and reaches $\phi\simeq 10^6 eV$ far away.

It is notable that the limiting value of $\Delta{m}^2$ is extremely close to the LMA (large mixing angle)solution of solar neutrino data\cite{Anomalies-cpt}; regarding the missing angle, the vanishing of $<\phi>$ and the resulting suppression of mixing in the dense regions (as was discussed in the previous section) suggests that the dominant share in the flavour change comes from non-oscillatory probability ${\sin\theta_{12}}^2$.  The matching angle is similar to those of LMA solution, i.e. $\theta \sim 33^{\circ}$\cite{Anomalies-cpt}.


As an outlook, one can study whether $\phi$ is the quintessence dark energy. In our model, during the cosmological evolution, $\phi$ becomes operative when $\rho_m<\rho_c$, while for $\rho_m>\rho_c$,  $\phi$ is freeze out at $\phi=0$. For example at the big bang nucleosynthesis  (BBN), where $z\sim 10^8$ ($a\sim 10^{-8}$), we have  $\rho_m^{BBN}\sim 10^{24}M_P^2H_0^2\sim 10^{13}eV^4$,  and if  $\rho_c<10^{13} eV^4$ the scalar field settles at $\phi=0$ in this stage. The requirement that $\phi$ be responsible for the late time Universe acceleration,  without affecting BBN, structure formation, and so on, may lead us to assume that the symmetry breaking has occurred recently.
In the symmetron model, this assumption implies a large mass for the quintessence, giving rise only to a short temporary acceleration \cite{mat}, which is inconsistent with astrophysical data. This leads people to add a cosmological constant term to the potential (like $V_0$ in (\ref{6})), which makes the model similar to $\Lambda CDM$ in cosmological scale.
 In our model, if one insists to relate the present acceleration of the Universe to a symmetry breaking which happened in the present epoch, then he must take $\epsilon R_0\sim \mu^2$, where $R_0$ is the present Ricci scalar. Nowadays, the Universe is dominated by dark energy $\rho_d$, and cold dark matter, such that  $\rho_d\simeq \frac{7}{3} \rho_m$ and $w_d\simeq -1$, where $w_d$ is the dark energy equation of state parameter \cite{planck}.  Hence $31\epsilon \Omega_m\sim \frac{\mu^2}{H_0^2}$, where $\Omega_m=\frac{\rho_m}{3M_P^2 H_0^2}\simeq 0.3$. $H_0$ is the present Hubble parameter. Collecting all together, we deduce that if the present acceleration is due to the symmetry breaking in the present era, then $\frac{\mu^2}{H_0^2}\sim 3\epsilon$. This is similar to the result obtained in \cite{kho} (by replacing $\epsilon$ with $\frac{M_P^2}{M^2}$).  But in the nonminimally coupled quintessence model (\ref{1}),   the Universe may experience both (positive)acceleration and super-acceleration phases \cite{nm1,nm2}. The acceleration occurs when\cite{nm1}
\begin{equation}\label{f1}
\left(1-\frac{\epsilon \phi^2}{M_P^2}\right)\rho_m-2V+\left(\frac{1}{2}-3\epsilon\right)\dot{\phi}^2+3\epsilon^2 R\phi^2+3\epsilon \phi \frac{dV}{d\phi}<0.
\end{equation}
For the potential (\ref{6}), and before the symmetry breaking, i.e. when the matter is very dense, we have $\phi=0$,  and (\ref{f1}) reduces to $\rho_m-2V_0<0$.  Hence for $V_0>0$, the acceleration may happen before the symmetry breaking. If we take $V_0=0$, the acceleration does not occur in this stage, and the acceleration of the Universe requires that $\rho_m<\rho_c$. But how long after $\rho_m=\rho_c$ the acceleration begins, and how long continues,  depends on the parameters of the model specially on nonminimal coupling of $\phi$ which affects its evolution \cite{sad2}.

In the end, it is worth to note that, we do not expect to meet instabilities arisen in neutrino dark energy in the cosmological extension of our model. In neutrino-dark energy models, when neutrinos become non-relativistic, the shape of the effective potential changes and the quintessence follows adiabatically the minimum of this potential, giving rise to the Universe acceleration. The adiabaticity in the non relativistic regime of neutrinos results in instabilities and formation of neutrinos nuggets \cite{afsh}. These instabilities do not occur if one discards adiabatic evolution \cite{bean}.  Our model is quite different from mass varying neutrino-dark energy model: the change of the quintessence potential, and consequently the symmetry breaking is due to the matter density (baryonic matter within or outside the astrophysical objects or dark matter in cosmological scales) and not to the behavior of the neutrinos, so we can evade from the constraints required by the neutrino-dark energy model such as the adiabaticity and so on.


\begin{thebibliography}{99}
\bibitem{osc1} B. Pontecorvo, Zh. Eksp. Teor. Fiz, 34, 247 (1958).
\bibitem{osc2} S.P. Mikheyev, A.Y. Smirnov, Nuovo Cimento C 9, 17 (1986).
\bibitem{osc3} L. Wolfenstein, phys. Rev. D 18, 958 (1978).
\bibitem{osc4}M. C. Gonzalez-Garcia, M. Maltoni,  Phys. Rept 460, 1 (2008).
\bibitem{osc5} S. Chakraborty,  JCAP 10, 019 (2015).
\bibitem{osc6}S. Capozziello, G. Lambiase, Mod. Phys. Lett.A 14, 2193 (1999).
\bibitem{osc7} M. V. Chizhov, S. T. Petcov, Phys. Rev. D 63, 073003 (2001). 
\bibitem{osc8}M. V. Chizhov, S. T. Petcov, Phys. Rev. Lett. 83, 1096 (1999).
\bibitem{Anomalies-cpt} A.Yu. Smirnov, arXiv:1609.02386v2[hep-ph].
\bibitem{Anomalies-cpt2} M. Wurm, Phys. Rept 685, 1 (2017).
\bibitem{Anomalies-LSND} LSND collaboration, Nucl. Instrum. Meth. A  388, 149 (1997).
\bibitem{micro} A. A. Aguilar-Arevalo \emph{et al}, Phys. Rev. Lett. 98, 231801 (2007).
\bibitem{micro1} A. A. Aguilar-Arevalo  \emph{et al}. (MiniBooNE Collaboration), Phys. Rev. Lett. 102, 101802 (2009).
\bibitem{Proposed} T. Katori, V. A. Kostelecký, and R. Tayloe, Phys. Rev. D 74, 105009  (2006).
\bibitem{Proposed1} A. Kostelecky, M. Mewes, Phys. Rev. D 70, 031902  (2004).
\bibitem{Proposed2}V. Barger, D. Marfatia, and K. Whisnant, Phys. Lett. B 653, 267 (2007).
\bibitem{Proposed3} A. Strumia, Phys. Lett. B 539, 91 (2002).
\bibitem{sterile} A. Slosar, Phys. Rev. Lett 97, 041301 (2006).
\bibitem{sterile1} Daya Bay \& MINOS Collaborations, Phys. Rev. Lett 117, 151801 (2016).
\bibitem{sterile2} L. Feng, Jing-Fei Zhang, and X. Zhan, Eur. Phys. J. C  77, 418 (2017).
\bibitem{matter effects} L. Wolfenstein, Phys. Rev. D 17, 2369 (1978).
\bibitem{matter effects1} L. Wolfenstein, Phys. Rev. D 20, 2634 (1979).
\bibitem{matter effects2} S. P. Mikheyev, A. Yu. Smirnov, Sov. J. Nucl. Phys. 42, 913 (1985).
\bibitem{ns1}O. G. Miranda, H. Nunokawa,  New Journal of Physics 17(9), 095002 (2015).
\bibitem{ns2}Y. Farzan, M. Tortola  arXiv:1710.09360v1 [hep-ph].
\bibitem{gold}G. J. Stephenson Jr., T. Goldman, B. H. J. McKellar, 	Mod. Phys. Lett. A 12, 2391 (1997).
\bibitem{string}A. Halprin, C. N. Leung, Phys. Lett. B 416, 361 (1998).
\bibitem{first} D. B. Kaplan, A. E. Nelson, and N. Weiner, Phys. Rev. Lett 93, 091801 (2004).
\bibitem{first1}V. Barger, P. Huber, and Danny Marfatia, Phys. Rev. Lett 95, 211802 (2005).
\bibitem{MaVaN} R. Fardon, A. E. Nelson, and N. Weiner,  JCAP 04110, 005 (2004).
\bibitem{sad} H. M. Sadjadi, V. Anari, 	Phys. Rev. D 95, 123521 (2017).
\bibitem{on} R. Onofrio, Phys. Rev. D 86, 087501 (2012).
\bibitem{Screening} K. A. Olive, M. Pospelov, Phys. Rev. D 77, 043524 (2008).
\bibitem{Screening1} K. Hinterbichler, J. Khoury, Phys. Rev. Lett 104, 231301 (2010).
\bibitem{mat}H. M. Sadjadi, M. Honardoost, and H. R. Sepangi, Phys. Dark Univ. 14, 40 (2016).
\bibitem{mat1}M. Honardoost, H. M. Sadjadi, and H. R. Sepangi, Gen. Rel. Grav. 48, 125 (2016),  arXiv:1508.06022 [gr-qc].
\bibitem{sad2}H. M. Sadjadi,  JCAP 01, 031 (2017), arXiv:1609.04292 [gr-qc] .
\bibitem{sad4}H. M. Sadjadi,  Phys. Rev. D 92, 123538 (2015).
\bibitem{mot}M. Honardoost, D. F. Mota, and H. R. Sepangi,  JCAP 11 (2017) 018, arXiv:1704.02572 [gr-qc].
\bibitem{Our}G. J. Jr. Stephenson \emph{et al}, Int. J. Mod. Phys. A 13, 2765 (1998).
\bibitem{Our1}N. F. Bell, E. Pierpaoli, and K. Sigurdson, Phy. Rev. D 73, 063523 (2005).
\bibitem{Our2} P. S. Pasquini, O. L. G. Peres, Phy. Rev. D 93, 053007 (2017).
\bibitem{Our4} A. Berlin, Phys. Rev. Lett 117, 231801 (2016).
\bibitem{wein} S. Weinberg, \emph{Gravitation and Cosmology: Principles and Applications of the General Theory of Relativity }, John Wiley \& Sons, Inc. (1972).
\bibitem{eq}L. Hui, A. Nicolis, and Ch. W. Stubbs, Phys. Rev. D 80, 104002 (2009).
\bibitem{eq1}C. Burrage, J. Sakstein, arXiv:1709.09071 [astro-ph.CO], 2017; J. Sakstein, arXiv:1710.03156 [astro-ph.CO], 2017.
\bibitem{Xi}X. Bi, B. Feng, H. Li, and X. Zhang, Phys.Rev. D 72, 123523 (2005).
\bibitem{cl}G. Cleaver, M. Cvetic, J. R. Espinosa, L. Everett, and P. Langacker. Phys. Rev. D 57, 2701 (1998.
\bibitem{phd}E. J. Weinberg, Ph.D. thesis, arXiv:hep-th/0507214.
\bibitem{fab}M. Fabbrichesi, A. Urbano, Phys. Rev. D 92, 015028 (2015).
\bibitem{Bu}G. Bambhaniya, P. S. Bhupal Dev, S. Goswami, S., and W. Rodejohann, Phys. Rev. D 95, 095016 (2017); F. Vissani,
Phys. Rev. D 57, 7027 (1998).
\bibitem{msnp}Z. Maki  M. Nakagawa, and S. Sakata , Prog. Theor. Phys. 28, 870 (1962).
\bibitem{bible} N. Fornengo, C. Giunti, C. W. Kim, and J. Song, Phy. Rev. D 56, 1895 (1997).
\bibitem{density}M. Cirelli, M.C. Gonzalez-Garcia, and C. Pena-Garay, Nucl. Phys. B 719, 219 (2005).
\bibitem{sun} D. D. Clayton, \emph{Principles of Stellar Evolution and Nucleosynthesis}, Chicago University Press, Reprint edition (1984).
\bibitem{dsolar}N. P. Pitjev and E. V. Pitjeva, Astronomy Letters 39, 141 (2013);
arXiv:1306.5534 [astro-ph.EP].
\bibitem{nm1}V. Faraoni, Phys. Rev. D 62, 023504 (2000).
\bibitem{nm2}C. Feng, X. Li, and E. N. Saridakis, Phys. Rev. D 82, 023526 (2010);  K. Nozari, S. D. Sadatian, Phys. Lett. B 676, 1 (2009);   P. Wang, P. Wu, and H. Yu,  Eur. Phys. J. C, 72, 2245 (2012); Chao-Qiang Geng, Chung-Chi Lee, Yi-Peng Wu,  arXiv:1512.04019v2 [astro-ph.CO]; F. Cicciarella, M. Pieroni, arXiv:1611.10074 [gr-qc]; M. Sharif, I. Nawazish, Eur. Phys. J. C 77, 198 (2017); C. Gomes, J. G. Rosa, and O. Bertolami,  arXiv:1611.02124 [gr-qc];
S. Bahamonde, M. Marciu, and P. Rudra, arXiv:1802.09155v1 [gr-qc].
\bibitem{planck}P. A. R. Ade et al. (Planck Collaboration), Astron. Astrophys.
594, A13 (2016).
\bibitem{kho}K. Hinterbichler, J. Khoury, A. Levy, and A.  Matas, Phys. Rev. D 84, 103521 (2011).
\bibitem{afsh} N. Afshordi, M. Zaldarriaga, and K. Kohri,  Phys. Rev. D 72, 065024 (2005).
\bibitem{bean}R. Bean, E. E. Flanagan, and M. Trodden, Phys. Rev. D 78, 023009 (2008).
\end{thebibliography}
\end{document}